\documentclass[pdflatex,sn-mathphys-num]{sn-jnl}

\usepackage{graphicx}%
\usepackage{multirow}%
\usepackage{amsmath,amssymb,amsfonts}%
\usepackage{amsthm}%
\usepackage{mathrsfs}%
\usepackage[title]{appendix}%
\usepackage{xcolor}%
\usepackage{textcomp}%
\usepackage{manyfoot}%
\usepackage{booktabs}%
\usepackage{algorithm}%
\usepackage{algorithmicx}%
\usepackage{algpseudocode}%
\usepackage{listings}%
\usepackage{upquote}

\begin{document}

\title[Validating and monitoring bibliographic and citation data in OpenCitations collections]{Validating and monitoring bibliographic and citation data in OpenCitations collections}

\author[1,2]{\fnm{Ivan} \sur{Heibi}}\email{ivan.heibi2@unibo.it}
\equalcont{These authors contributed equally to this work.}

\author[1,2]{\fnm{Silvio} \sur{Peroni}}\email{silvio.peroni@unibo.it}
\equalcont{These authors contributed equally to this work.}

\author[1]{\fnm{Elia} \sur{Rizzetto}}\email{elia.rizzetto2@unibo.it}
\equalcont{These authors contributed equally to this work.}

\affil[1]{\orgdiv{Research Centre for Open Scholarly Metadata}, \orgname{Department of Classical Philology and Italian Studies, University of Bologna}, \orgaddress{\street{Via Zamboni, 32}, \city{Bologna (BO)}, \postcode{40126}, \country{Italy}}}

\affil[2]{\orgdiv{Digital Humanities Advanced Research Centre (/DH.arc)}, \orgname{Department of Classical Philology and Italian Studies, University of Bologna}, \orgaddress{\street{Via Zamboni, 32}, \city{Bologna (BO)}, \postcode{40126}, \country{Italy}}}

\abstract{\textbf{Purpose:}
The increasing emphasis on data quantity in research infrastructures has highlighted the need for equally robust mechanisms ensuring data quality, particularly in bibliographic and citation datasets. This paper addresses the challenge of maintaining high-quality open research information within OpenCitations, a community-guided Open Science Infrastructure, by introducing tools for validating and monitoring bibliographic metadata and citation data.

\textbf{Methods:}
We developed a custom validation tool tailored to the OpenCitations Data Model (OCDM), designed to detect and explain ingestion errors from heterogeneous sources, whether due to upstream data inconsistencies or internal software bugs. Additionally, a quality monitoring tool was created to track known data issues post-publication. These tools were applied in two scenarios: (1) validating metadata and citations from Matilda, a potential future source, and (2) monitoring data quality in the existing OpenCitations Meta dataset.

\textbf{Results:}
The validation tool successfully identified a variety of structural and semantic issues in the Matilda dataset, demonstrating its precision. The monitoring tool enabled the detection of recurring problems in the OpenCitations Meta collection, as well as their quantification. Together, these tools proved effective in enhancing the reliability of OpenCitations’ published data.

\textbf{Conclusion:}
The presented validation and monitoring tools represent a step toward ensuring high-quality bibliographic data in open research infrastructures, though they are limited to the data model adopted by OpenCitations. Future developments are aimed at expanding to additional data sources, with particular regard to crowdsourced data.}


\maketitle

\section{Introduction}\label{sec1}

In recent years, significant focus and effort have been spent on gathering as much data as possible \textemdash a period we, as scholars, have rephrased as the Big Data era \cite{chenBigDataSurvey2014}. These dynamics have characterised several scholarly disciplines, research and industrial contexts. We have been involved in retrieving and organising as much data as possible, sure that the available quantity of such information is one of the most prominent and essential characteristics we should care of when working with data-intensive processes, thus preferring big volumes of data over small samples of good quality information.

Such a view of preferring quantity over quality is a debated theme in several contexts, recently including research assessment exercises. Indeed, several initiatives around this topic, such as the Coalition for Advancing Research Assessment (CoARA, https://coara.eu/) \cite{coalitionforadvancingresearchassessmentAgreementReformingResearch2022}, have pushed a lot for the use of quality over quantity in research assessment, leaving the quantitative dimension as a possibility for supporting responsibly peer-reviewed evaluation.

However, the focus on quality concerns the evaluation processes and the ground data used to inform them. Even when considering the responsible use of quantitative indicators in this context, there is a clear push for adopting open research information \textendash see the Barcelona Declaration on Open Research Information (DORI, https://barcelona-declaration.org/) \cite{barcelonadeclarationonopenresearchinformationBarcelonaDeclarationOpen2024} \textendash that, in particular in the context of research assessment, enables to inform appropriately transparent and high-quality assessment processes using ``highest quality data" \cite{hicksBibliometricsLeidenManifesto2015}. Thus, the open infrastructures involved in this context, i.e. shared digital research infrastructures designed to support data-driven science while adhering to FAIR (Findable, Accessible, Interoperable, Reusable) principles \cite{wilkinsonFAIRGuidingPrinciples2016a}, have the responsibility to ensure high-quality and reliable data validated via automated and manual checks entailed by adopting robust data management practices \cite{manolaOpenInfrastructuresResponsible2025}. All these aspects have been considered for years in the ingestion processes and quality checks introduced and implemented at OpenCitations.

OpenCitations is a non-for-profit and community-guided Open Science Infrastructure compliant with the Principles for Open Scholarly Infrastructures (POSI, https://openscholarlyinfrastructure.org/) \cite{bilderPrinciplesOpenScholarly2020}, dedicated to the gathering and publication of open bibliographic metadata and citation data, made available as complete dumps in several formats and programmatically accessible via appropriate REST APIs. OpenCitations provides two main collections: OpenCitations Index \cite{heibiOpenCitationsIndexDescription2024}, storing citations links between scholarly bibliographic resources, and OpenCitations Meta \cite{massariOpenCitationsMeta2024}, storing the basic bibliographic metadata (title, authors, year of publication, publication venue, publishers, identifiers) of the citing and cited entities involved in the citations available in the OpenCitations Index. These datasets are derived from diverse sources \textendash which currently (as of 14 April 2025) include Crossref \cite{hendricksCrossrefSustainableSource2020a}, DataCite \cite{datacitemetadataworkinggroupDataCiteMetadataSchema2024}, NIH Open Citation Collection \cite{hutchinsNIHOpenCitation2019}, OpenAIRE \cite{atzoriOpenAIREWorkflowsData2017, manghiOpenAIREplusEuropeanScholarly2012}, and the Japan Link Centre \cite{katoJapanLinkCenter2012} \textendash and such primary data are reshaped according to the OpenCitations Data Model (OCDM) \cite{daquinoOpenCitationsDataModel2020} via an established ingestion workflow \cite{heibiOpenCitationsIndexDescription2024, malinekOpenBibliographicalData2024}.

However, ingesting large volumes of data from heterogeneous sources introduces potential errors and inconsistencies, which we should identify and explain to devise processes to correct them and, thus, improve the quality of the data. These errors fall into two broad categories:

\begin{itemize}
    \item errors in primary sources, which may arise from human inaccuracies or software bugs in the originating systems;
    \item issues in the OpenCitations ingestion software due to bugs in the conversion and ingestion processes implemented within the infrastructure.
\end{itemize}

The implemented workflow devised at OpenCitations automatically addresses many of the problems arising from the ingestion process, either by sanitising invalid data or discarding them where automatic correction is not feasible. However, there still remain challenges that must be solved by fixing bugs in the core components of the ingestion process (to avoid its repetition) and by correcting the current problems in the published data either by developing and running ad hoc computational tools or by curating the identified issues manually using appropriate interfaces (e.g. HERITRACE \cite{massariHERITRACETracingEvolution2024}).

A preliminary validation process and post-publication monitoring tool are essential to identify and understand incorrect data, minimise information loss during ingestion, and ensure the accuracy of the final datasets. To meet these objectives, and by extending the preliminary insights introduced in \cite{peroniToolValidatingMonitoring2025}, OpenCitations has developed a custom validation tool presented in this article, which is designed to provide the precision and granularity necessary for addressing the unique requirements of the OCDM. This validation tool is accompanied by a tool for monitoring the quality of data published in the two OpenCitations collections to track the existence of known errors in the data and help implement effective curation, prevention and correction strategies. We show the application of both tools to check the validity of another source that OpenCitations will consider in the future, i.e. Matilda \cite{tornyMatildaBuildingBibliographic2019}, and to identify and monitor existing issues in the data available in OpenCitations Meta.

The rest of this paper is structured as follows. Section ``Material and Methods" analyses the case study and the data features involved and describes the methodology followed to develop viable solutions for validating bibliographic metadata and citation data and monitoring their quality. Section ``Implementation” provides more technical details on the implementation of the software components. Section “Applicative Scenarios” illustrates the application of the implemented software considering two use cases, citations and metadata provided by Matilda (for data validation) and the data currently in OpenCitations Meta (for quality monitoring). Section ``Related Works" discusses other related work about the quality assessment of RDF data. Finally, Section “Conclusions” sums up the overall work and sketches out some future developments.

\section{Material and Methods}\label{sec:methods}

This section presents an overview of OpenCitations' ingestion workflow and describes the proposed methodology for pre-ingestion data validation and post-ingestion quality assurance.

\subsection{The data and the current ingestion workflow}\label{subs:methods:data}

In the OpenCitations Index , citations are represented as first-class data entities, meaning that each citation is represented as an entity in its own right, representing a directed link between two other entities (\textit{publication A} cites \textit{publication B}) with its properties, including: citing entity, cited entity, citation creation date, and citation timespan (i.e. the difference in days between the date of publication of the citing entity and the date of publication of the cited entity). All the data in OpenCitations Index is collected from raw citation data openly provided by other external sources, and it is then published under a CC0 waiver. The current (as of 14 April 2025) sources are Crossref \cite{hendricksCrossrefSustainableSource2020a}, DataCite \cite{datacitemetadataworkinggroupDataCiteMetadataSchema2024}, the National Institute of Health Open Citation Collection (NIH-OCC) \cite{hutchinsNIHOpenCitation2019}, OpenAIRE \cite{atzoriOpenAIREWorkflowsData2017}, and the Japan Link Center (JaLC) \cite{katoJapanLinkCenter2012}. The metadata of the publications involved as either citing or cited entities in the OpenCitations Index are stored in OpenCitations Meta \cite{massariOpenCitationsMeta2024}. For each publication, this collection provides details including their persistent identifiers (PIDs) for the publication (e.g. DOI), its title, the publication type (e.g. journal article, book, dataset, etc.), the publication date, the venue and its PIDs, the page interval, the issue and volume numbers, and the name and PIDs of the agents involved in the publication, i.e. authors, editors and publisher.

Gathering citation data and bibliographic metadata from diverse primary sources and unifying them into the OpenCitations Index and the OpenCitations Meta poses significant challenges since each source represents the data in its own way, and some information might overlap or differ across the sources. To overcome this challenge, a workflow has been developed to reshape the gathered data according to the OpenCitations Data Model (OCDM) \cite{daquinoOpenCitationsDataModel2020} and then ingest it into the collections. The OCDM is a data model built by reusing existing ontologies for describing information in the scholarly bibliographic domain and essentially consists, in the scope of data validation and error detection, of the fundamental set of rules defining the correct relationships and properties of all entities in OpenCitations Meta and OpenCitations Index.
The workflow currently implemented for making bibliographic metadata and citation data coming from external sources OCDM-compliant and ingesting it into OpenCitations’ collections consists of three steps \cite{heibiOpenCitationsIndexDescription2024}:

\begin{enumerate}
    \item \textit{\textbf{Source Preprocess}}. This step reshapes the data by implementing a metadata crosswalk from the diverse data models used by original sources to the OCDM, managing the differences in information content, structure and representation. A central operation in this phase is the normalisation and validation of external PIDs, such as DOIs for publications or ORCIDs for authors and editors. 
    
    The output of the software dedicated to this step, the OpenCitations Data Sources Converter \cite{morettiOc_ds_converter2024}, are two OCDM-compliant tables which are used in the two following steps: one storing the bibliographic metadata for each publication involved in a citation (where each row represents a publication and columns store the values for supported metadata content), the other storing citations (where each row represents a citation and two columns store the PIDs of the citing and the cited publication, whose scheme depend on the original source, e.g. citations from Crossref are represented as DOI-to-DOI citations).
    
    \item \textit{\textbf{Meta Process}}. This step populates the OpenCitations Meta collection. Starting from the table produced in the first step, the bibliographic metadata to ingest is automatically curated by deduplicating records (i.e. table rows) that feature the same PID as another record and normalising and correcting their values. Records representing entities that had been registered in OpenCitations Meta in previous ingestions can be used to enrich the already available metadata for those entities or to merge them into a single entity (in the case the external PIDs appearing to pertain to a single entity in the record are instead linked to separate entities in OpenCitations Meta). Each entity in OpenCitations Meta is represented by the OpenCitations Meta Identifier (OMID), a PID that is minted and assigned to the entity in the moment of its generation: this is a crucial feature, as it allows the following step of the workflow to uniquely identify the publications linked by citations without relying on external identifiers.
    
    The output of this step, whose complete methodology and implementation are detailed thoroughly in \cite{massariOpenCitationsMeta2024}, is the OpenCitations Meta dataset itself, stored both in a database and as dump files. Notably, the software responsible for the operations mentioned above also generates and stores in RDF files provenance information for each entity, keeping track of the agent that created, modified, merged or deleted it, the time of the action and the primary source providing the data.
    \item \textit{\textbf{Index Process}}. This step processes the citation tables from the Source Preprocess phase where each citation is represented as a link between two external PIDs (e.g., DOI-to-DOI, PMID-to-PMID). By making use of a mapping between these external identifiers and the OMID of the entity they have been associated with in the previous step (Meta Process), it converts these links into OMID-to-OMID citations, each of which is uniquely identified as a first-class entity by an Open Citation Identifier (OCI). Similarly to step 2, the process output consists of the OpenCitations Index dataset, with citation data stored in a database and as dump files, and provenance information saved in files only.
\end{enumerate}

The workflow briefly described above is currently only applied to data from authoritative sources, such as Crossref or DataCite, which structure their data according to a defined data model. While this workflow is undoubtedly useful \textendash particularly because it allows for the ingestion of a large volume of data with each execution \textendash its application is effectively limited to data sources where implementing the metadata crosswalk from the source data model to the OCDM (as outlined in the Source Preprocess step) is feasible or advantageous. Since this process requires notable effort to manage the idiosyncratic complexities of each source, creating a custom data conversion system for each source may not be an applicable strategy, especially for sources that cannot provide certain conditions that facilitate this process (e.g. a defined data model, clear documentation of the data structure, etc.). Nonetheless, there are organisations and individuals who, despite lacking these characteristics, hold high-quality bibliographic data that is not yet easily accessible or reusable. Ingesting this data into collections like those of OpenCitations is crucial to making a large number of up-to-date citations and bibliographic metadata openly available.

As has been already pointed out by OpenCitations \cite{heibiCrowdsourcingOpenCitations2019}, an effective solution to broaden the number of open scholarly bibliographic data can be represented by crowdsourcing the data itself: users would be able to directly submit, via a dedicated service, tables containing citations and metadata to be ingested into OpenCitations Index and OpenCitations Meta respectively. Users would need to submit tables that are already OCDM-compliant and formatted to be natively processed by the relevant OpenCitations software (i.e. equivalent to the output of the current workflow’s first step, Source Preprocess); in creating them, they should follow the guide provided in two reference documents \cite{massariHowStructureCitations2022, massariHowProduceWellformed2022}, in order for these tables to be interpreted correctly in the Meta Process and Index Process steps of the workflow. As the tabular documents obtained this way would not be built via a controlled internal process, validating them becomes imperative to ensure data quality. To fulfil this need, we have developed a custom validation tool described in the following subsection.

\subsection{Pre-ingestion Validation}\label{subs:methods:validation}

The tabular format has been chosen for user submissions in that it is approachable even by scholars, researchers and professionals with little coding skills, yet the inherent complexity of the relationships and information expressable with the OCDM can fit into such a format only following precise rules, defined in \cite{massariHowStructureCitations2022, massariHowProduceWellformed2022}. Following the nomenclature in these reference documents, we will henceforth refer to the table storing metadata as META-CSV and to the table storing citations as CITS-CSV.

In META-CSV, each row represents a bibliographic resource, i.e. a publication, and the eleven columns specify: the identifiers associated with the resource; the title; the surname, name, and identifiers of its authors and editors; the publication date; the venue (i.e. another bibliographic resource containing the represented document, e.g. the journal containing the article represented in the row); the volume of the venue containing the document; the issue of the venue containing the document; the page range; the type of publication; and the name and identifiers of the publisher.

In CITS-CSV, each row represents one citation, and the four columns store the values for the identifiers and the publication date of the citing and the cited bibliographic resource.
META-CSV and CITS-CSV tables have a layered structure that adds complexity to their validation. Beyond their tabular organisation of rows and columns, field values within each cell can consist of either single data units or collections of multiple data units separated by specific delimiters. These individual units, termed “items”, represent the minimal “portion” used by the document to define a specific piece of information and, therefore, must be validated individually. For example, in CITS-CSV, the identifier fields for citing and cited resources may contain multiple items, while in META-CSV, fields for identifiers, authors, venues, publishers, and editors may similarly admit multiple items.

\begin{figure}
  \centering
  \includegraphics[width=\linewidth]{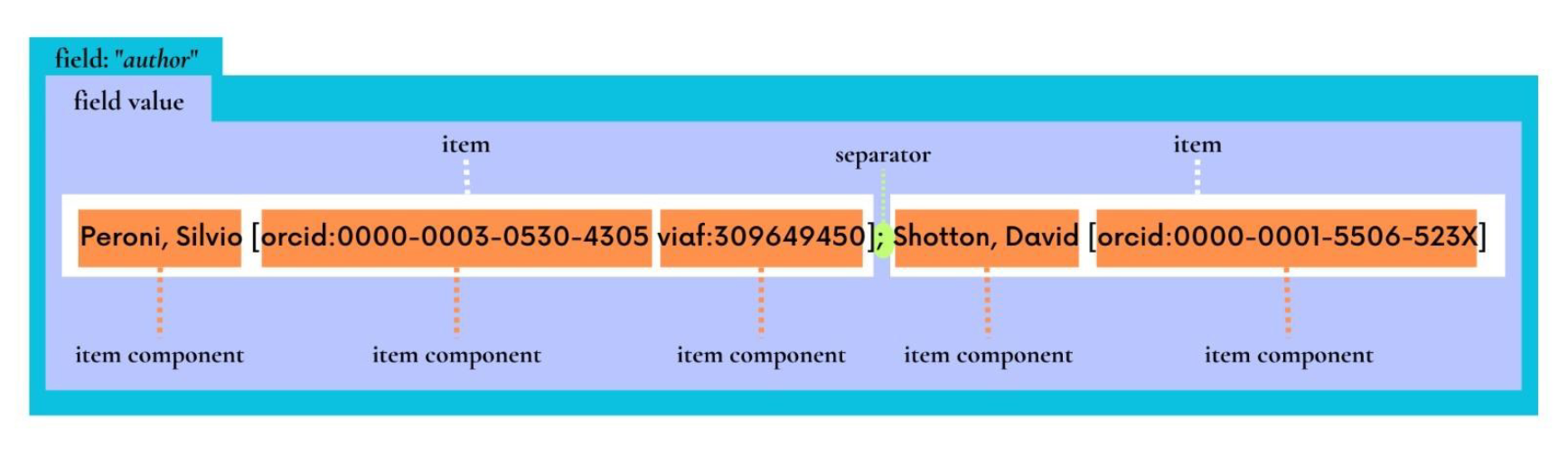}
  \caption{The abstract representation of the internal structure of the table cell containing the data for the author of a bibliographic resource (see Table \ref{tab:meta}). The cell contains two items, each of which corresponds to the entity of an author; each items has internal components of different kinds (the plain text of the surname and name, and the series of identifiers).}
  \label{fig:author}
\end{figure}

\begin{figure}
  \centering
  \includegraphics[scale=0.25]{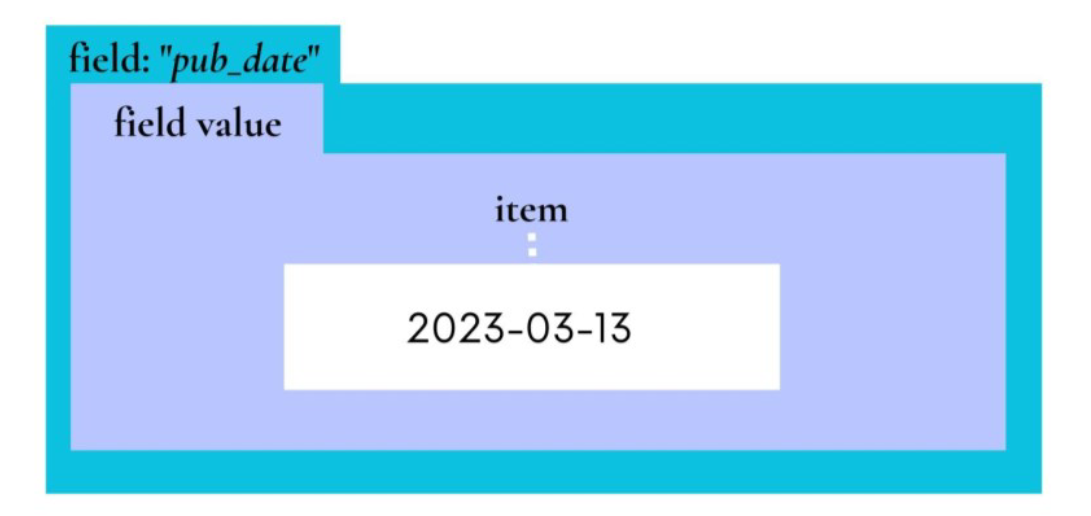}
  \caption{The abstract representation of the internal structure of the table cell containing the data for the publication date of a bibliography resource (see Table \ref{tab:meta}). The cell contains only one item, which corresponds to the value of the publication date. The publication date field always has a value containing a single item.}
  \label{fig:pubdate}
\end{figure}

Adding to this complexity, in META-CSV, some fields contain items composed of smaller components, such as names and identifiers of entities (e.g., authors, editors, publishers, and venues). Each component requires distinct validation rules based on its type, leading to diverse validation requirements for the content of a single field. Table \ref{tab:meta}, Figure \ref{fig:pubdate}, and Figure \ref{fig:author} can be used to understand the abstract representation of the structure of the table.

\begin{table*}
  \caption{Two sample table cells of META-CSV, storing the surnames, names and identifiers of the authors of a bibliographic resource and the publication date of the bibliographic resource.}
  \label{tab:meta}
  \begin{tabular}{|c|p{0.7\textwidth}|l|c|}
    \toprule
    ...& author& pub\_date & ...\\
    \midrule
    ... & Peroni, Silvio [orcid:0000-0003-0530-4305 viaf:309649450]; Shotton, David [orcid:0000-0051-5506-523X]& 2023-03-13& ...\\
  \bottomrule
\end{tabular}
\end{table*}

Validation rules for the tables encompass both formatting/syntactic and content criteria. Syntactic rules are defined in the specifications to write well-formed documents \cite{massariHowProduceWellformed2022, massariHowStructureCitations2022} and ensure proper data types, formats, and required fields. The other rules extend beyond syntax, addressing requirements such as the existence of referenced identifiers in relevant registries and the correctness of the relationships expressed in the table, also for those requirements that are not explicitly mentioned in the table specifications (e.g. a META-CSV row corresponding to a bibliographic resource to which a given type has been assigned may have only a certain set of values in the identifier field to be compliant with OCDM). 

At the beginning of the validator designing phase, we first identified all the applicable validation rules for each of the two documents and grouped them into four different categories: rules related to the format and syntax of the document as prescribed by OpenCitations; rules based on the externally-defined syntax of PIDs (e.g. the valid structure of a DOI value); rules verifying the existence of an entity in the real world; and rules checking the relationships between the values. These categories have been used to structure the validation process into four levels, applied sequentially to the table elements:

\begin{enumerate}
    \item \textit{Wellformedness}. This step ensures the document complies with the syntactic rules defined in \cite{massariHowProduceWellformed2022, massariHowStructureCitations2022} to generate well-formed tables, e.g. supported identifier schemes, correct date formats, etc. Errors at this level block further validation of affected items.
    \item \textit{ID Syntax}. All PID values are checked against syntax rules defined by their issuing organizations, ensuring formats such as the ones for DOI\footnote{https://www.doi.org/}, ORCID\footnote{https://orcid.org/}, and PMID\footnote{https://pubmed.ncbi.nlm.nih.gov/} are correctly applied.
    \item \textit{ID Existence}. The existence of mentioned entities in the real world is verified by using their associated identifiers as a proxy: PIDs are queried against official databases to confirm they are actually registered as such.
    \item \textit{Semantics}. Applied only to META-CSV, this step verifies the consistency of the relationships between data points (e.g. verifying that the PID associated with the bibliographic resource represented in a row is compatible with the resource type).
\end{enumerate}

The implementation of the validation process has been guided by the following design principles:

\begin{enumerate}
    \item \textit{Maximum granularity}. Each document is validated by applying checks on its smallest parts (in most cases items, but if applicable also sub-parts of an item) to maximise granularity in the output and identify faulty table elements with high specificity.
    \item \textit{Maximum coverage at each execution}. At each execution of the process, the entire table is validated from start to end, i.e. without stopping the process if an error is found: all detectable errors are collected during the process and returned as a comprehensive collection in the output validation report. This process makes it easier for users to correct errors, since they are enabled to potentially address all issues in one correction cycle, avoiding repeated submissions.
    \item \textit{Non-redundancy}. A single item that has already failed a check is not validated against the rest of the rules, as it will need to be modified by the user. This principle only applies in cases where compliance with one rule is a prerequisite for compliance with the other rules for the same item. Otherwise, i.e. if the outcomes of two checks on the same item are mutually independent, both checks are executed straight away (i.e. before the user intervenes with any corrections).
\end{enumerate}

The custom validator has been designed with the aim of providing a precise and information-rich feedback on the validity status of the data that is both suitable for programmatic use and human-readable. Machine-readability is essential for the primary objective of discarding invalid data before ingestion automatically, and for granting the possibility to use the validation output in other applications (e.g. for double-checking internally generated documents in the Source Preprocess phase of the ingestion workflow). Human-readability and user-friendliness are key to the fulfilment of the other fundamental objective of the validator: providing users with a tool to better understand how to create correct bibliodata tables.

The output is provided as a report listing all detected errors from a single execution of the validation process. Each error includes:

\begin{itemize}
    \item \textit{Position details}: the exact location of all the single pieces of data involved in that error (relative to the whole document) and whether the error regards a single item, multiple fields, or multiple rows.
    \item \textit{Validation level}: The validation level where the error occurred.
    \item \textit{Type of issue}: Whether the issue is a blocking ``error'' or a non-blocking ``warning''.
    \item \textit{Unique error label}: A short label indicating the category to which each error instance belongs, which can be used by a machine to process the output.
    \item \textit{User message}: Natural language explanation of the error and its potential causes.
\end{itemize}

Particular attention has been given to finding a feasible way to express the position of the error in the document, so that the specific data points involved in the error could be retrieved and processed automatically, while at the same time grant the user the possibility to exactly see the single parts of the document involved in the error. Nonetheless, the format specially designed to indicate error positions reflects the complexity of the internal structure of the table and can be cumbersome for humans to read. To solve this limitation, we went further in the direction of user-friendliness and paired the validator with a component entirely dedicated to the visualisation of the validation report in a graphical interface. This solution allows users to grasp the basic information about the errors more easily, visualising directly, on a new tabular representation of the input document, where they are located and their explanation.

\subsection{Post-ingestion data quality assurance}
\label{subs:methods:monitoring}

Managing large collections of bibliographic and citation metadata from diverse sources requires a robust and systematic process to ensure data quality. Despite preventive measures such as validation during ingestion, errors can persist or emerge over time. To maintain data integrity, it is essential to monitor the correctness of ingested data regularly, implement correction strategies, and evaluate the success of these interventions.

A data monitoring tool has been developed for OpenCitations Meta and OpenCitations Index to address this need. This tool systematically searches for and tracks the presence of pre-identified and categorised errors in the data, providing actionable insights to guide corrections and assess the outcomes of previous improvements.

As a preliminary step, all known errors existing in OpenCitations Meta and OpenCitations Index have been collected and described via the use of a basic framework, which enables us to keep track, for each error, of details like the assigned error label, actual examples, a description of the issue, the interested collection, and any strategy or programming solution to find actual data affected by the error.

From here, a subset of error types has been selected, including those reproducible by querying the databases of the two collections, which are accessible online via the SPARQL endpoint. This is because retrieving content by querying the databases for a given pattern is much faster than other solutions, like the programmatic analysis of the whole dump. 

The monitoring process works by a simple logic: for each error, the data are tested by trying to retrieve from the associated collection any results that fall within the pattern representing it; if any result is retrieved, it means that there is wrong data (i.e. not compatible with OCDM or at any rate not expected) and the test for that particular error fails. The outcome of all the tests (i.e. whether it passed or not) is gathered during the process and returned in the final output of the monitor, which includes:
\begin{itemize}
    \item general details about the execution of the process: the SPARQL endpoint URL of the accessed database, the queried collection (i.e. either OpenCitations Meta or OpenCitations Index), the date and time of the execution, the total running time, the path of the configuration file used to specify which errors have been tested.
    \item the details for each single error: label, natural language description, retrieving method (i.e. the exact SPARQL query), test result and details about the execution of that single test, including its individual running time and execution errors in case they were raised.
\end{itemize}

The monitoring tool has been designed with extendability, automation, and accessibility in mind. Its modular architecture ensures that new error tests can be added seamlessly, provided the errors can be represented through SPARQL queries. This flexibility allows the system to adapt continuously as new types of errors are discovered and defined.

To facilitate continuous oversight, the monitor can operate fully automatically, with periodic, scheduled executions. Currently, the process runs every Monday for both OpenCitations Meta and OpenCitations Index. 
Similarly to the validator, the monitor prioritizes user-friendliness by presenting its results in both machine-readable and human-readable formats. This dual representation supports diverse use cases, from automated integration into workflows to manual examination of findings. Notably, as part of the automated workflow, the human-readable results are used to update a publicly accessible web page, offering full transparency into the latest monitoring outcomes. The results can be viewed at: https://ocmonitor.opencitations.net/.

\section{Implementation}
\label{sec:implementation}

This section presents technical details on the implementation of the methodology described above.

\subsection{oc\_validator}
\label{subs:impl:validator}

The validator tool described in Section \ref{sec:methods} has been implemented in a Python software named \verb|oc_validator|, available as a public repository\footnote{https://github.com/opencitations/oc\_validator} under an ISC license and as an installable library in PyPI\footnote{https://pypi.org/project/oc-validator/}.

\begin{figure}
  \centering
  \includegraphics[width=\linewidth]{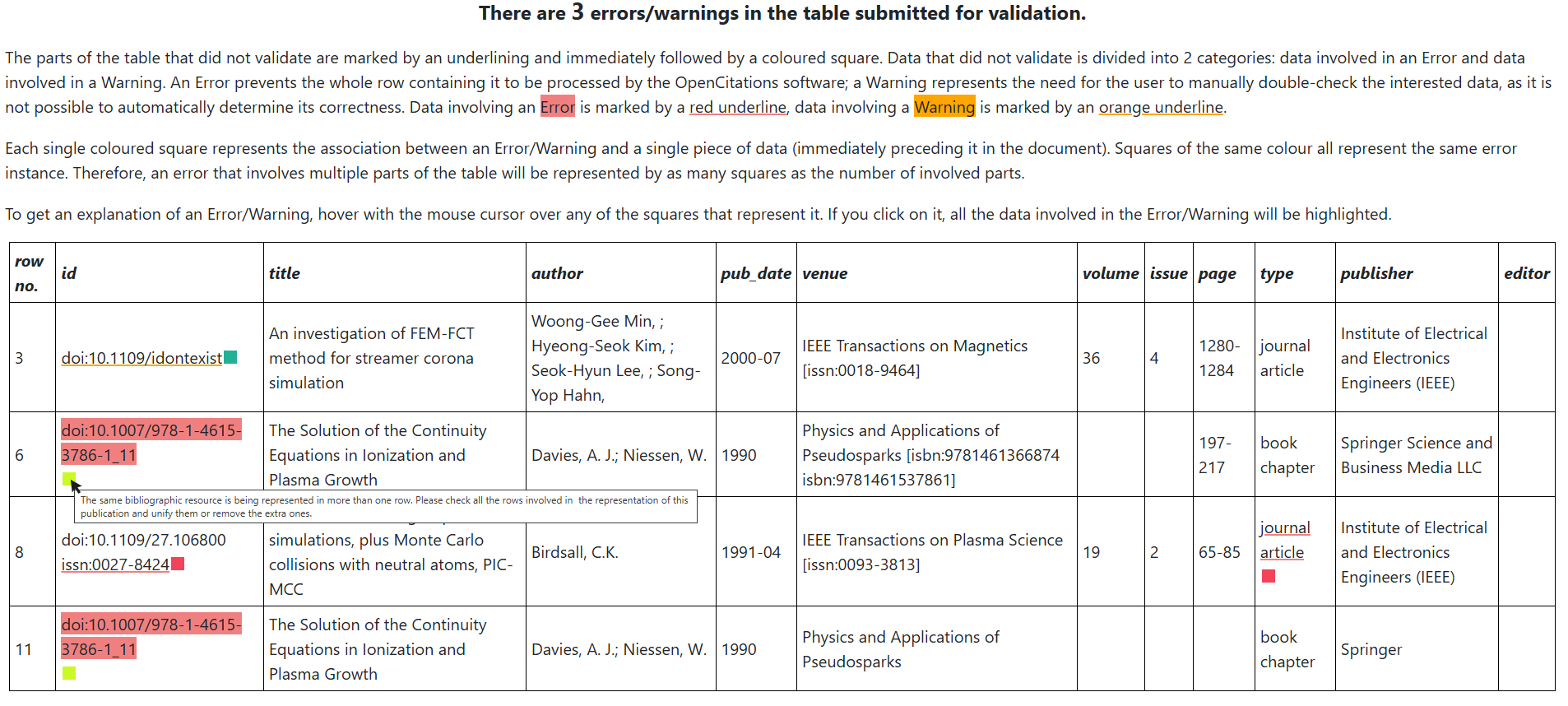}
  \caption{A screenshot of an example HTML page for the visualisation of a META-CSV document validation report.}
  \label{fig:gui-row}
\end{figure}

The main process is managed by a dedicated class, named \verb|Validator|, that takes as input the path to the table to validate (a CSV file that must be formatted as either META-CSV or CITS-CSV) and the path to the directory where the output files will be stored. As this interface only deals with one document at a time, another class was added to enable the simultaneous cross-validation of both metadata and citation data contained in two separate documents (also formatted as  META-CSV and CITS-CSV respectively), which simply wraps in two instances of \verb|Validator|, one for either document type.

As a first step of the process, the type of table is automatically determined, and a specific method managing the related operations is called accordingly. Internally, in fact, the processes for validating META-CSV and CITS-CSV are distinct, though their inner workings are similar. 

The validation process then iterates over all the table rows and columns (i.e. fields) and, depending on the field name, elaborates the internal content of the cell (which is initially interpreted as a single string) according to its abstract internal structure. The appropriate validation rules are sequentially checked by executing specific functions for each of the extracted items and their sub-components. Each of the four validation levels has its corresponding class, and each validation rule of the level is checked by a dedicated class method. In compliance with the design principle of non-redundancy mentioned in the Material and Methods section, if an element fails a check it is not tested for the other validation rules that might apply to it, unless the latter are independent of the first check's outcome.

Whenever a check fails, a dictionary object is generated representing the error and added to a list that will constitute the validation process output. An example of an error object is represented below in the form of a JSON object:

\begin{lstlisting}
{
    "validation_level": "csv_wellformedness",
    "error_type": "error",
    "error_label": "duplicate_br",
    "valid": false,
    "message": "The same bibliographic resource is being represented in more than one row. Please check all the rows involved in  the representation of this publication and unify them or remove the extra ones.",
    "position": {
        "located_in": "row",
        "table": {
            "2": {"id": [0, 1]},
            "3": {"id": [0, 1]}
        }
    }
}    
\end{lstlisting}

When the end of the document is reached, and all the elements of the table have been validated, the list containing the error object is stored in a JSON file constituting the machine-readable output of the process and is converted into a TXT file summarising the latter, which consists of its human-readable version.

After the validation phase, the JSON output creates a graphical user interface as an HTML page to visualise the validation outcome better. In the HTML document, the integrated CSS styling and JavaScript code allow users to interact with the content. The rows of the original document that contain one or more errors are presented in an HTML table, where the location of the error is signalled by underlining exactly the wrong content associated with it and accompanying it with a square. By clicking on each square, the related faulty content is highlighted in all the involved locations, while hovering on it with the cursor shows the explanation of the error. An example the interface is provided as a static image in Figure \ref{fig:gui-row}.

\subsection{oc\_monitor}

\begin{figure}[b!]
  \centering
  \includegraphics[width=\linewidth]{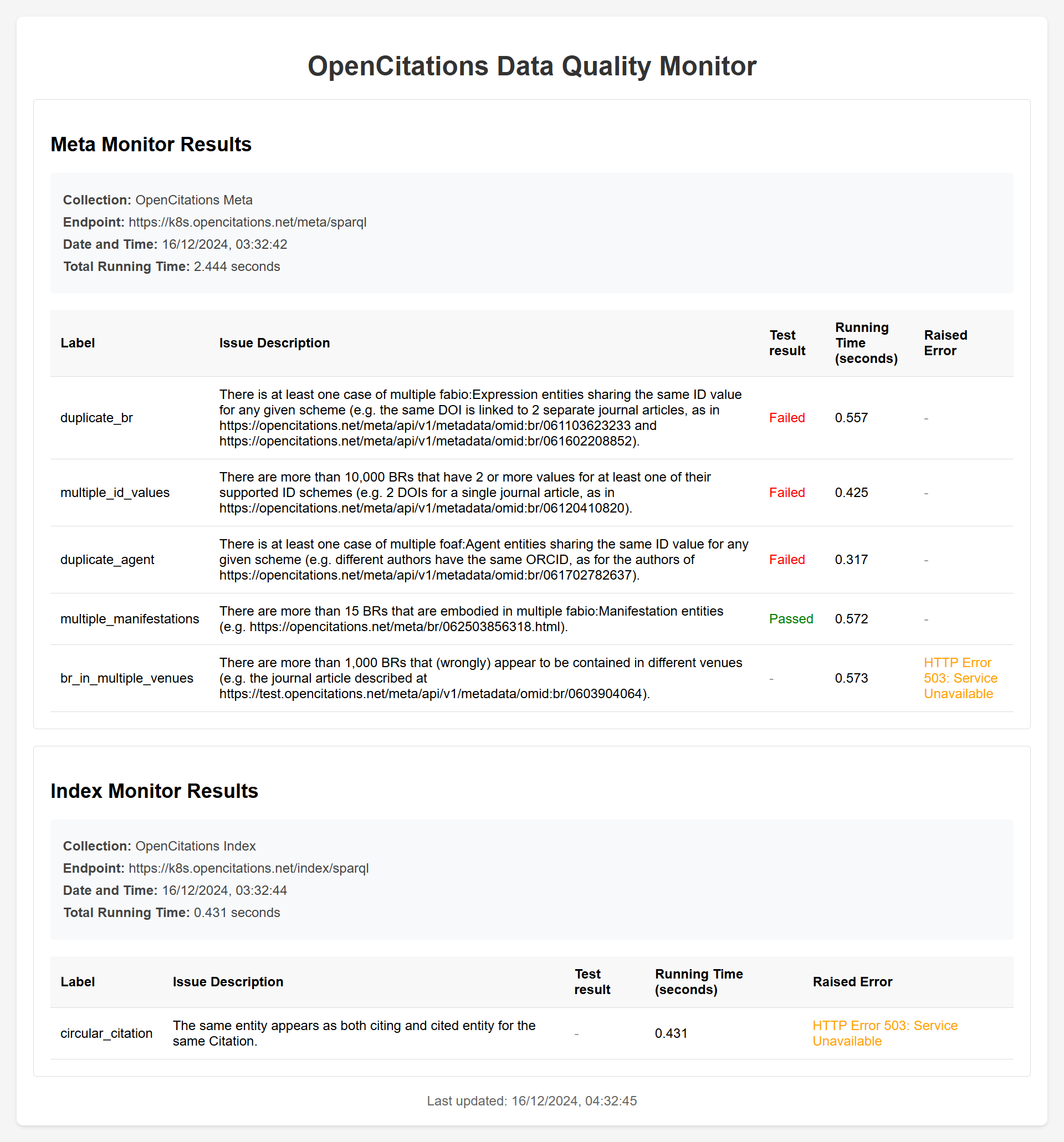}
  \caption{A screenshot of an example HTML page showing the results of oc\_monitor. The test results and the execution time values are for illustrative purposes only.}
  \label{fig:monitor}
\end{figure}

The monitor tool has also been implemented in the Python software \verb|oc_monitor|, available as a public repository\footnote{https://github.com/opencitations/oc\_monitor} under an ISC license and as an installable library\footnote{https://pypi.org/project/oc-monitor/} in PyPI.

\label{subs:impl:monitor}
The central component in the monitor software is the JSON configuration file that is passed as argument of the two separate classes that actually manage the process: \verb|MetaMonitor|, in charge of looking for errors in OpenCitations Meta, and \verb|IndexMonitor|, looking for them in OpenCitations Index. The configuration file specifies the URL of the SPARQL endpoint to interrogate and a list of objects representing an error or an issue in the data. Each of these objects contains: a short label for the issue; a textual description explaining the nature of the problem and possibly providing examples of faulty data; a flag to indicate whether the specific issue should be verified at the execution of the monitoring process; and the actual test for the error, a SPARQL query defining the pattern that would catch results if the error was present in the collection.

\verb|MetaMonitor| and \verb|IndexMonitor| must each be passed the appropriate configuration file to work properly, but their inner working is similar: for each test in the configuration file, the SPARQL query is executed against the specified endpoint and a report for the test is generated in the form of a dictionary. This object contains the information specified in the configuration (label, description and query text) and a boolean indicating whether the test passed or not. Moreover, details concerning the execution are included: running time of the query\footnote{The execution of each test requires from less than a second to less than 5 minutes, depending on the type of test and factors such as other ongoing processes on the working server, simultaneous network traffic, etc. Especially considering the small number of tests, using the SPARQL endpoint is much faster than processing the whole dump, currently comprising more than 4,8 billion triples for OpenCitations Meta and more than 9 billion triples for OpenCitations Index.} and, in case execution errors were raised, e.g. due to HTTP errors, the problem itself is reported too, providing also a useful insight on the status of the system infrastructure at the moment of the attempted connection to the server. An example of an object representing a test result is given below:

\begin{lstlisting}
    {
    "label": "duplicate_br",
    "description": "There is at least one case of multiple fabio:Expression entities sharing the same ID value for any given scheme (e.g. the same DOI is linked to 2 separate journal articles, as in omid:br/061103623233 and omid:br/061602208852).",
    "query": "PREFIX datacite: <http://purl.org/spar/datacite/>\nPREFIX literal: <http://www.essepuntato.it/2010/06/literalreification/>\nPREFIX fabio: <http://purl.org/spar/fabio/>\n\nASK {\n    ?br1 datacite:hasIdentifier/literal:hasLiteralValue ?lit ;\n    a fabio:Expression .\n    ?br2 datacite:hasIdentifier/literal:hasLiteralValue ?lit ;\n    a fabio:Expression .\n    FILTER(?br1 != ?br2)\n}",
    "run": {
        "got_result": true,
        "running_time": 2.4839844703674316,
        "error": null
    },
    "passed": false
}
\end{lstlisting}

The output of the monitoring phase consists of these objects, stored in a JSON file with some general details such as the exact date and time of the execution and the total running time.

As for the validator, the JSON output is used to create a more user-friendly visualisation of the monitoring outcome, in this case simply by restructuring the content into an HTML page. 

Notably, OpenCitations has paired oc\_monitor with an automatisation pipeline developed via GitHub Actions\footnote{https://github.com/features/actions} and available in the software's repository: every Monday, the monitor is run against both collections, and the latest results are added for storage in the public repository and exposed on a public web page\footnote{https://ocmonitor.opencitations.net/}. For an example of such HTML page see Figure \ref{fig:monitor}.

\section{Applicative Scenarios}
As a form of informal evaluation of the software, we focus on two distinct applicative scenarios introduced in two complementary parts. The first part focuses on the validation capabilities of the \textit{oc\_validator}, assessed through the analysis of small data dumps provided by Matilda, a bibliometric and bibliographic research tool developed in the context of open science \cite{tornyMatildaBuildingBibliographic2019}. The second part evaluates the system’s behaviour under realistic operational conditions by executing the \textit{oc\_monitor} on the OpenCitations Meta dataset. These two aspects provide a comprehensive assessment of both functional correctness and runtime performance. Details of each evaluation phase are presented in the following subsections and related material, including input data, results, and code, are publicly available in \cite{rizzettoMaterialEvaluationOc_validator2025}.

\subsection{oc\_validator \textendash Matilda case study}
\label{subs:eval-matilda}

Matilda \cite{tornyMatildaBuildingBibliographic2019} is a bibliometric and bibliographic research tool. It was designed to overcome the limitations of traditional proprietary systems by integrating a wide range of openly available data sources and supporting user-driven customization. Matilda provides a flexible infrastructure for bibliometric analysis, capable of accounting for diverse document types, languages, and access models. Its development reflects a broader shift toward openness, transparency, and accessibility in scholarly communication.

As part of an ongoing collaboration with OpenCitations, Matilda has contributed a collection of its bibliographic and citation data, dated September 2024, for integration into the OpenCitations Meta and OpenCitations Index, respectively. This contribution constitutes a concrete use case that is particularly well-suited for evaluation using the \textit{oc\_validator} tool introduced in this paper. The provided dataset of Matilda includes a sample of 3,464 citations, such that each citation is defined by the identifier of the citing and cited entity, and 5,101 bibliographic entities, such that each entity (row in the dataset) is accompanied by its corresponding metadata.

As described in Table \ref{tab:eval-matilda-metadata}, the validation process identified the most frequent warning concerning the nonexistent identifiers for bibliographic resources (br\_id\_existence, 849 instances), followed by logically inconsistent page intervals (page\_interval, 664 instances), suggesting that the starting page appears to be greater than the end page. A significant number of entries also presented a malformed page format (page\_format, 522), which was categorized as an error due to its direct impact on the consistency of the data. Additional issues included titles entirely in uppercase (uppercase\_title, 60), and a less frequent but structurally critical errors involved duplicate entries of responsible agents (duplicate\_ra, 23), incorrect formatting of agent representations (people\_item\_format, 14), and malformed bibliographic resource identifiers (br\_id\_format, 1). Although relatively limited, these final results still contain errors that require attention. Lastly, a small number of cases involved unregistered responsible agent identifiers (ra\_id\_existence, 5), indicating a need for closer control over external identifier validation.

On the other hand, when analysing the citations provided by Matilda – Table \ref{tab:eval-matilda-citations}, the results identified a limited number of issues related to identifier validity. The most common warning was the presence of unregistered bibliographic resource identifiers (br\_id\_existence, 296 cases), followed by self-citations (self\_citation, 61 cases), which may be legitimate but the rarity of these cases might suggest a further verification over the given data. Only one instance of a malformed identifier (br\_id\_format) was found, indicating strong adherence to syntax conventions by the Matilda sample.

These findings on the Matilda data, underscore the effectiveness of the OC Validator, particularly in identifying both high-impact errors and softer formatting issues, thereby supporting the goal of improving metadata quality and consistency in scholarly datasets.

\begin{sidewaystable*}
\caption{The results generated by the \textit{oc\_validator}, detailing the type, description, and frequency of errors and warnings identified in the metadata of the bibliographic resources of Matilda.}\label{tab:eval-matilda-metadata}
\begin{tabular}{l l p{0.5\textwidth} r}
\toprule
error\_label & error\_type & message & \# \\
\midrule
page\_format & Error & The value of 'page' is not well-formed. There must always be a starting page, followed by an hyphen, followed by the end page ... & 522 \\
duplicate\_ra & Error & The same responsible agent (author/editor/publisher) is reported more than once within the same cell. Please remove the extra occurrence(s). & 23 \\
people\_item\_format & Error & The value representing the responsible agent entity is not well-formed. The entity for a responsible agent is represented by the name of the person/organization, followed by a single whitespace and one or more associated identifiers, … & 14 \\
br\_id\_format & Error & The value in this field is not expressed in compliance with the syntax of OpenCitations CITS-CSV/META-CSV. Each identifier in 'citing\_id'/'cited\_id' ... & 1 \\
br\_id\_existence & Warning & The ID is not registered anywhere as a persistent identifier for a bibliographic resource, i.e. it does not exist. & 849 \\
page\_interval & Warning & The specified page interval seems to be impossible: the start page appears to be greater than the end page. & 664 \\
uppercase\_title & Warning & The whole title of the publication is uppercase. Are you sure? Please double-check the actual title of the publication. & 60 \\
ra\_id\_existence & Warning & The ID is not registered as a persistent identifier for any bibliographic resource, i.e. it does not exist. & 5 \\
\botrule
\end{tabular}
\end{sidewaystable*}

\begin{table*}[h]
\caption{The results generated by the oc\_validator for citation data of Matilda, showing the types, descriptions, and occurrences of errors and warnings detected.}\label{tab:eval-matilda-citations}
\begin{tabular}{l l p{0.54\textwidth} r}
\toprule
error\_label & error\_type & message & \# \\
\midrule
br\_id\_format & Error & The value in this field is not expressed in compliance with the syntax of OpenCitations CITS-CSV/META-CSV. Each identifier in 'citing\_id'/'cited\_id' … & 1 \\
br\_id\_existence & Warning & The ID is not registered anywhere as a persistent identifier for a bibliographic resource & 296 \\
self\_citation & Warning & It seems that a circular citation is being represented: the bibliographic resource appears to be citing itself. & 61 \\
\botrule
\end{tabular}
\end{table*}

\subsection{oc\_monitor \textendash OpenCitations Meta}
\label{subs:eval-monitor}
Starting from the queries used in the automated workflow for monitoring data quality issues in OpenCitations Meta \textendash which are designed to return a boolean indicating the presence or absence of an error at the time of execution \textendash we developed a new set of queries aimed at quantifying these previously identified issues. Specifically, for each issue represented by a query, we retrieved the total number of affected resources in the dataset.

The results of this analysis, conducted using the live triplestore data as of 11 April 2025, are presented in Table \ref{tab:eval-monitor}. The table includes only those tests that detected the presence of actual data issues, reporting the corresponding number of resources affected for each type of error.

\begin{table*}[h!]
\caption{The results of oc\_monitor configured with custom SPARQL queries over the OpenCitations Meta SPARQL endpoint. For each error there are a label, a description in natural language, and the total count of distinct entities affected by the error.}\label{tab:eval-monitor}
\begin{tabular}{l p{0.41\textwidth} r}
\toprule
Label & Issue Description & \# \\
\midrule
duplicate\_br\_count & Bibliographic resources sharing the same ID value for any given scheme (e.g. the same DOI is linked to 2 separate journal articles). & 1,388,761 \\
duplicate\_agent\_count & Agents sharing the same ID value for any given scheme (e.g. different authors have the same ORCID). & 2,544,914 \\
duplicate\_id\_count & Identifiers (datacite:Identifier entities) sharing the same literal value for the same scheme. & 1,388,761 \\
multiple\_manifestations\_count & Bibliographic resources appearing to be embodied in multiple fabio:Manifestations entities. & 13 \\
br\_in\_multiple\_venues\_count & Bibliographic resources appearing to be contained in different venues. & 760,011 \\
br\_with\_multiple\_id\_values\_count & There are bibliographic resources that have 2 or more values for at least one of their supported ID schemes (e.g. 2 DOIs for a single journal article). & 760,011 \\
\botrule
\end{tabular}
\end{table*}

These results reveal that, in most cases, the number of resources affected by each issue is too large to be addressed manually. Nevertheless, these errors still represent a relatively small proportion of the overall dataset. For instance, although there are 1,388,761 bibliographic resources in OpenCitations Meta that share a persistent identifier with at least one other resource (as indicated by the duplicate\_br\_count label), this accounts for only 1.1\% of the total 121,302,680 bibliographic resources present in the collection at the time of analysis.
Similarly, the issue labelled duplicate\_agent\_count indicates that 2,544,914 agent entities (such as authors, editors, or publishers) share a persistent identifier with at least one other agent. Despite the large absolute number, these cases represent just 0.8\% of the 333,356,609 total agent entities in the dataset.
Overall, this quantitative perspective helps to contextualize the severity and prevalence of different types of data quality issues in OpenCitations Meta, supporting more informed decisions regarding prioritization and potential remediation strategies — whether automated or manual.

\section{Related Work}
\label{sec:related-works}
As concerns RDF data quality assessment, there have been several endeavours to tackle this problem \cite{zaveriQualityAssessmentLinked2015}, though the great majority of the associated tools (where they exist) either require a high degree of manual configuration or provide information of limited use. An interesting work in this field, whose approach resembles the one proposed by our paper, is presented in \cite{kontokostasTestdrivenEvaluationLinked2014}. Here the authors present a methodology and a tool for assessing the quality of RDF data by following a test-driven approach inspired by software engineering. They propose the concept of Data Quality Test Patterns (DQTPs), which encapsulate common data quality issues into structured SPARQL query templates. These patterns are then used to instantiate actual test-cases (SPARQL queries) by binding specific values to the variables in the templates. Notably, test-cases can be automatically generated with Test Auto Generators (TAGs), which interrogate the RDF data to test and try to instantiate test-cases based on the OWL axioms and RDFS constructs defined in the ontologies used therein. With TAGs, it is possible to ensure that the data complies with simple schema constraints such as property domain, range and cardinality. For more complex tests or tests that are not derivable from the constraints formalised in the ontologies, the authors suggest to manually instantiate specific tests by re-using the DQTPs. The methodology presented in \cite{kontokostasTestdrivenEvaluationLinked2014} also defines coverage metrics, which measure the adequacy of the test cases by assessing how well they capture different aspects of data quality (e.g. property domain coverage, class membership), represents the test-cases in RDF and associates a URI to each of them.

While this methodology proves effective for datasets with broad and heterogeneous structures \textemdash such as DBpedia, which was evaluated by the authors \textemdash its applicability to more controlled environments like OpenCitations requires careful consideration. The schemas included in the ontologies reused by the OpenCitations Data Model (OCDM) are well-suited for verifying general or simpler semantic relationships, but they are often too basic to support the automatic generation of meaningful tests tailored to the specific constraints and application context relevant to OpenCitations. In contrast to crowdsourced datasets like DBpedia, the data in OpenCitations is converted and ingested within a strictly controlled environment, ensuring that many fundamental constraints are already satisfied by design. As a result, tests generated purely from ontology schemas would, in many cases, verify relationships whose correctness is already guaranteed, adding little value to the quality assessment process.

Moreover, many of the properties and classes defined in the ontologies reused by OpenCitations are not currently utilized in its datasets. This means that an automatic test generation approach, as described by \cite{kontokostasTestdrivenEvaluationLinked2014}, would likely produce a significant number of non-applicable test cases. Furthermore, OpenCitations' current post-ingestion quality assessment relies on a small set of specific tests, carefully designed based on known issues that have been previously identified in the dataset. At this stage, implementing a more generalized pattern-based testing methodology would require substantial effort without offering clear advantages. Instantiating tests manually provides greater flexibility, allowing for fine-grained selection of the rules to be checked while at the same time avoiding the computational cost of executing unnecessary test cases: coherently with the aim of providing an insight on the data quality that is a good representation of its fitness for use, our approach focuses on ensuring that quality assessments can be performed frequently and that the results are presented in a way that is clear and easily interpretable.

Additionally, there are practical concerns regarding the long-term viability of adopting the methodology proposed by \cite{kontokostasTestdrivenEvaluationLinked2014}, as the software tool implementing their approach does not appear to be actively maintained.

Nevertheless, \cite{kontokostasTestdrivenEvaluationLinked2014} remains relevant for large-scale RDF quality assessment, and some of its aspects could be valuable for OpenCitations in the future. The scalability of its approach makes it suitable for analyzing extensive datasets, though, as mentioned, in the case of OpenCitations the results might include information of limited relevance. The representation of test cases in RDF with structured metadata is another strength of their methodology, offering a formalized approach to documenting and managing quality assessments. Furthermore, relying on a library of patterns facilitates the creation of new test-cases in a systematic and reusable way: in case the number of pre-identified issues should grow, it might be an interesting avenue to explore also for OpenCitations Meta and OpenCitations Index.

\section{Conclusions}
\label{sec:conclusion}

This work has presented a framework for ensuring the quality of bibliographic and citation data within the OpenCitations Meta and OpenCitations Index. By addressing pre-ingestion validation, we have built a system aimed at verifying the compliance of the data eligible for ingestion with the data model adopted by OpenCitations and with its existing ingestion workflow: the implemented validator ensures the syntactic and semantic correctness of documents storing data to ingest, identifying errors with granular precision and enabling both programmatic and user-friendly use. As concerns post-ingestion quality monitoring, we developed a tool that can be used to continuously evaluate the current data by verifying its status for known, previously identified issues.

A key feature of both the validation and the monitoring tools is their human-oriented design: the endeavour to provide both internal and external users with clear and accessible information led to pairing the machine-readable results of both tools with human-readable and user-friendly interfaces.

While these tools can facilitate data quality management, certain limitations remain. The validator is inherently tied to OpenCitations' specific table formats (META-CSV and CITS-CSV) and the rules of the OCDM, making it unsuitable for use with other data structures and models. Moreover, the monitor relies on SPARQL queries to detect errors, which limits its capacity to identify issues that manifest only in data dumps or API behaviours. 

An evaluation of both \textit{oc\_validator} and \textit{oc\_monitor} was conducted. The \textit{oc\_validator} was assessed using the Matilda dataset, allowing for a detailed analysis of metadata and citation data quality issues. On the other hand, \textit{oc\_monitor} was evaluated in the context of OpenCitations Meta, providing insights into its effectiveness in tracking the status and availability of large-scale bibliographic data. These evaluations support the reliability and applicability of the tools in real-world scholarly infrastructures.

Future efforts will focus on integrating the validator into a fully automated data submission workflow, allowing seamless validation during the ingestion process. This objective is particularly relevant within the context of data crowdsourcing, as it would streamline operations that can be done automatically while at the same time valuing the role of human agents in ensuring data quality.

\section{Acknowledgments}
    This project has been partially funded by the European Union's Horizon Europe framework programme under grant agreement No. 101095129 (GraspOS Project).

\bibliography{bibliography}

\end{document}